\documentclass[useAMS,usegraphicx]{mn2e}
\usepackage{times}
\usepackage{amsmath,amssymb}
\usepackage{color}

\title
[Testing a new feedback model]
{
Galaxy luminosity function and its cosmological evolution: Testing a new feedback model depending on galaxy-scale dust opacity
}

\author
[R. Makiya et al.]
{
R.~Makiya, $^1$\thanks{Email: makiya@ioa.s.u-tokyo.ac.jp}
T. ~Totani, $^2$
M.~A.~R.~Kobayashi, $^3$
M.~Nagashima, $^4$
and T.~T.~Takeuchi $^5$
\\
$^1$ Institute of Astronomy, The University of Tokyo, Mitaka, Tokyo 181-0015,
Japan\\
$^2$ Department of Astronomy, School of Science, The University of Tokyo, Hongo, Bunkyo-ku, Tokyo 113-0033\\
$^3$ Research Center for Space and Cosmic Evolution, Ehime University, Bunkyo-cho, Matsuyama 790-8577, Japan\\
$^4$ Faculty of Education, Nagasaki University, 1-14 Bunkyo-machi, Nagasaki 852-8521, Japan\\
$^5$ Institute for Advanced Research, Nagoya University, Furo-cho, Chikusa-ku, Nagoya 464-8601, Japan
}
\begin{document}

\voffset-.45in
\maketitle

\begin{abstract}
We present a new version of a semi-analytic model of cosmological
galaxy formation, incorporating a star formation law with a feedback
depending on the galaxy-scale mean dust opacity and metallicity,
motivated by recent observations of star formation in nearby
galaxies and theoretical considerations. This new model is used to
investigate the effect of such a feedback on shaping the galaxy
luminosity function and its evolution.  Star formation activity is
significantly suppressed in dwarf galaxies by the new feedback effect,
and the faint-end slope of local luminosity functions can be
reproduced with a reasonable strength of supernova feedback, which is in
contrast to the previous models that require a rather extreme strength
of supernova feedback. Our model can also reproduce the early
appearance of massive galaxies manifested in the bright-end of high
redshift $K$-band luminosity functions.  Though some of the previous
models also succeeded in reproducing this, they assumed a star formation
law depending on the galaxy-scale dynamical time, which is not
supported by observations.  We argue that the feedback depending on
dust opacity (or metal column density) is essential, rather than that
simply depending on gas column density, to get these results.
\end{abstract}


\section{Introduction}
\label{sec:intro}

The basic picture of galaxy formation and evolution in the
cosmological context can be explained in the standard $\Lambda$ cold
dark matter (CDM) cosmology. Particularly, large scale clustering
properties and formation and evolution of dark matter halos can
reliably be predicted by the theory of gravity.  However, in order to
obtain the full picture of cosmological galaxy formation, we must
solve complicated processes of baryonic physics, such as gas cooling,
star formation, feedback, galaxy mergers, and so on.  One of the key
observables about galaxies that must be explained by the theory of
cosmological galaxy formation is the luminosity functions (LFs) and their
evolution. Compared with the shape of dark matter halo mass function
predicted by the $\Lambda$CDM cosmology, the observed galaxy
LFs have two remarkable features: flatter faint-end
slopes and sharp exponential cut-off at the luminous/massive end (see
Benson et al. 2003 and references therein), which must be explained by
some baryonic processes.

A widely accepted solution to achieve a flat faint
end is supernova feedback, i.e., energy input into the interstellar medium
by supernova explosions to suppress star formation in small galaxies.
However, the problem is not yet completely solved at the quantitative
level. In fact, unreasonably high efficiency of supernova feedback to
remove cold interstellar gas in dwarf galaxies is necessary in many
existing theoretical models to reproduce the observed flat faint ends,
and such an extreme supernova feedback tends to produce discrepancies
with observations other than luminosity function shapes (Nagashima \&
Yoshii 2004, hereafter NY04; Nagashima et al. 2005; 
Bower et al. 2006, 2012; Guo et al. 2011;
Wang, Weinmann \& Neistein 2012; Mutch, Poole \& Croton 2013;
Puchwein \& Springel 2013; Hopkins et al. 2013).  
These results imply that another physical
effect may also be taking an important role to produce the observed
flat faint end slopes. 

For the massive end, a popular solution to suppress the formation of
too massive galaxies is the feedback by active galactic nuclei (AGNs;
e.g., Bower et al. 2006; Croton et al. 2006; Menci et al. 2008;
Somerville et al. 2008; Guo et al. 2011).  The AGN feedback can also
explain the observed trends of the early appearance of massive and
quiescent galaxies at high redshifts, and downsizing of star-forming
galaxies from high to low redshifts, which are apparently in
contradiction with the simple expectation in the $\Lambda$CDM
universe (e.g., Bower et al. 2006; Somerville et al. 2008).
However, there are large uncertainties about the physics of
AGN feedback both in theoretically and observationally. The current success
in explaining the observed trends by this process is based on rather
phenomenological modelings including highly uncertain parameters, and
further studies are required to confirm the quantitative influence of
this process on galaxy evolution.

Therefore it is still worth to explore yet other physical effects
working to shape galaxy LFs, which is the aim of this
paper. It is reasonable to expect that such an effect would be
manifested in the scaling laws about star formation efficiency.  The
relation between the surface densities of star formation rate (SFR) and gas
($\Sigma_{\rm SFR}$-$\Sigma_{\rm gas}$) has been a subject of
intensive research. It is popular to fit this relation by a power-law
(so-called Kennicutt-Schmidt law, Kennicutt 1998), but
recent observations indicate a cut-off around the total (i.e., {\hbox{H{\sc i} }} +
H$_2$) gas density of $\Sigma_{\rm gas} \sim 10\;M_\odot \ \rm
pc^{-2}$, under which SFR is suppressed and not well correlated with
gas density.  This threshold gas density for SFR can be interpreted
as a result of less efficient formation of cold molecular gas under
the threshold, while the star formation efficiency (SFE) from
molecular gas is rather universal in many different environments (Wong
\& Blitz 2002; Kennicutt et al. 2007; Bigiel et al. 2008, 2010; Leroy
et al. 2008; Blanc et al. 2009; Heiderman et al. 2010; Lada et
al. 2010, Schruba et al. 2011; see Kennicutt \& Evans 2012 and Schruba
2013 for reviews).

A likely physical origin of the suppression of H$_2$ formation under
the threshold is radiative feedback by UV photons produced by young
massive stars (Schaye 2004; Krumholz, Mckee \& Tumlinson 2008, 2009;
McKee \& Krumholz 2010; Hopkins et al. 2013).  The formation
of H$_2$ is driven by collisionally excited metal line cooling and
molecule formation on dust grain surfaces, which should be balanced
with molecule dissociations by UV photons and grain photoelectric
heating, both of which are energetically supplied by UV radiation
field. If a region in a galaxy is optically thick to UV radiation
field by dust grains, self-shielding of UV radiation would accelerate
H$_2$ formation. This implies that the more fundamental threshold
about star formation is not the total gas surface density but dust
opacity.  For a typical dust-to-gas ratio, the observationally
indicated threshold in $\Sigma_{\rm gas}$ is close to the value at
which the effective dust opacity $\tau_d^{\rm eff}$ becomes of order
unity, where $\tau_d^{\rm eff}$ is averaged over wavelength with a
weight of the heating radiation energy spectrum (Totani et al. 2011).

Therefore it is physically reasonable to expect that a galaxy-scale
mean value of $\tau_d^{\rm eff}$ has an important role in galaxy
formation and evolution.  A further observational support to this
picture comes from infrared observations. The relations between dust
temperature, galaxy size, and infrared luminosity of $\sim$1,000 nearby
star-forming galaxies indicate that almost all of them are in the
optically thick regime, and the distribution of dust opacity estimated
by gas-phase metal column density suddenly drops around $\tau_d^{\rm
  eff} \sim 1$, indicating less efficient formation of galaxies at
$\tau_d^{\rm eff} \la 1$ (Totani et al. 2011).

The purpose of this paper is to investigate the effect of the
radiative feedback depend on dust opacity, on cosmological galaxy
formation and evolution particularly about the shape of galaxy
luminosity functions.  The theory of structure formation in the
universe predicts that the mean surface density $M/r^2$ of dark halos
with mass $M$ and size $r$ nearly scales as $\propto M^{1/3} (1+z)^2$,
indicating higher gas surface density and dust opacity at higher
redshifts in more massive objects, and hence more efficient star
formation.  This may have a favorable effect to explain observations,
in a similar way to the feedbacks by supernovae and AGNs.

To investigate the effect quantitatively, we use a semi-analytic model
(SAM) of cosmological galaxy formation, {\it the Mitaka model}
(NY04). This is a model similar to general SAMs, in which formation
and evolution of dark matter halos are solved analytically or
calculated by N-body simulations, while complicated baryonic processes
are treated phenomenologically (for reviews, see Baugh 2006; Benson 2010).
In general, SAMs has many adjustable parameters and the effects of complicated physical processes on the LFs are degenerate (e.g., Neistein \& Weinmann 2010);
therefore a set of best-fit parameters may not be a quantitatively correct 
description of real galaxy formation. It should be noted that the most
important aim of this work is to examine the qualitative effects of 
the new feedback on luminosity functions. 

In most of the SAMs, the star formation rate is simply proportional to cold
gas mass, and the star formation time scale is modeled as a simple
function of the dynamical time scale of galaxy disks or DM halos (e.g.,
Cole et al. 2000; NY04).  Some models (e.g., Kauffmann 1996; Croton et
al. 2006; Somerville et al. 2008; Lagos et al. 2011; Wang et al. 2012)
incorporated the threshold of gas surface density below which star
formation activity is significantly suppressed.  In the models of
Kauffmann (1996), Croton et al. (2006), and Lagos et al. (2011), they
introduced the threshold of gas surface density motivated by the
Toomre stability criterion on a galactic scale (Toomre 1964).  In this
scenario the threshold of gas surface density increases with redshift,
and hence the threshold effect should be systematically different in
the cosmological context from the threshold by dust opacity considered
in this paper. Furthermore, some recent observations indicate that
star formation are controlled by the physical state of local
interstellar gas, rather than the dynamical state of an entire galaxy
(e.g., Leroy et al. 2008; Lada et al. 2010).

In other models, such as Somerville et al. (2008), a critical gas
surface density threshold for star formation is introduced motivated
from the observations of the $\Sigma_{\rm SFR}$-$\Sigma_{\rm gas}$
relation; however, to our knowledge there are no SAMs that consider a
feedback depending on dust surface density rather than gas density.
Recently Krumholz \& Dekel (2012) incorporated a star formation law
which depends on gas surface density and gas metallicity, and
discussed average evolution of typical galaxies without calculating
detailed merger histories of dark halos. The relation between the
luminosity function shapes and the dust opacity threshold of star
formation has not yet been discussed in previous studies.

This paper is organized as follows.  In section \ref{sec:model}, we
will describe our model particularly focusing on the modelings of star
formation and feedback.  In section \ref{sec:Results}, we show the
results of our model, and section \ref{sec:gasmodel} is devoted for
discussion.  We will summarize our work in section \ref{sec:summary}.
In this work, the cosmological parameters of $\Omega_{0}=0.3$,
$\Omega_{\Lambda}=0.7$, and $H_{0}=70~{\rm Mpc^{-1}~km~s^{-1}}$ are
adopted, and all magnitudes are expressed in the AB system.

\begin{figure*}
\begin{center}
    \begin{tabular}{cc}
		\includegraphics[width=0.49\hsize]{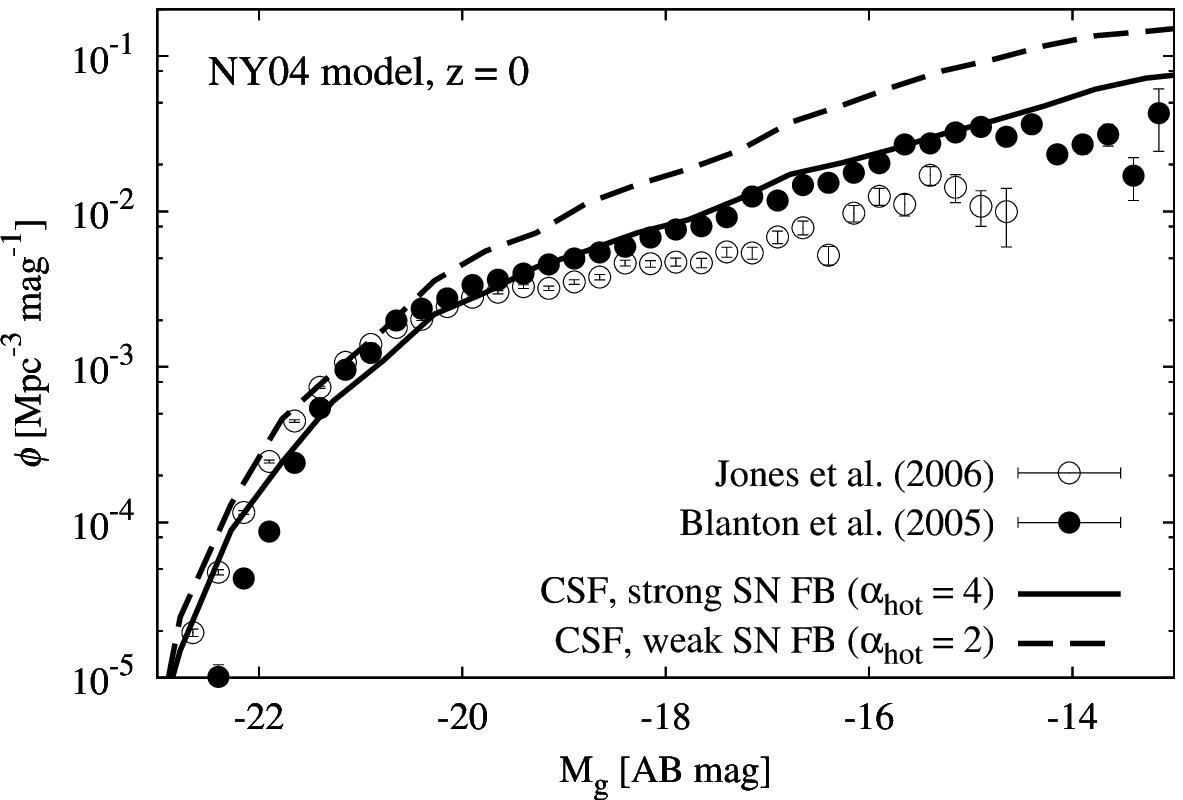} &
		\includegraphics[width=0.49\hsize]{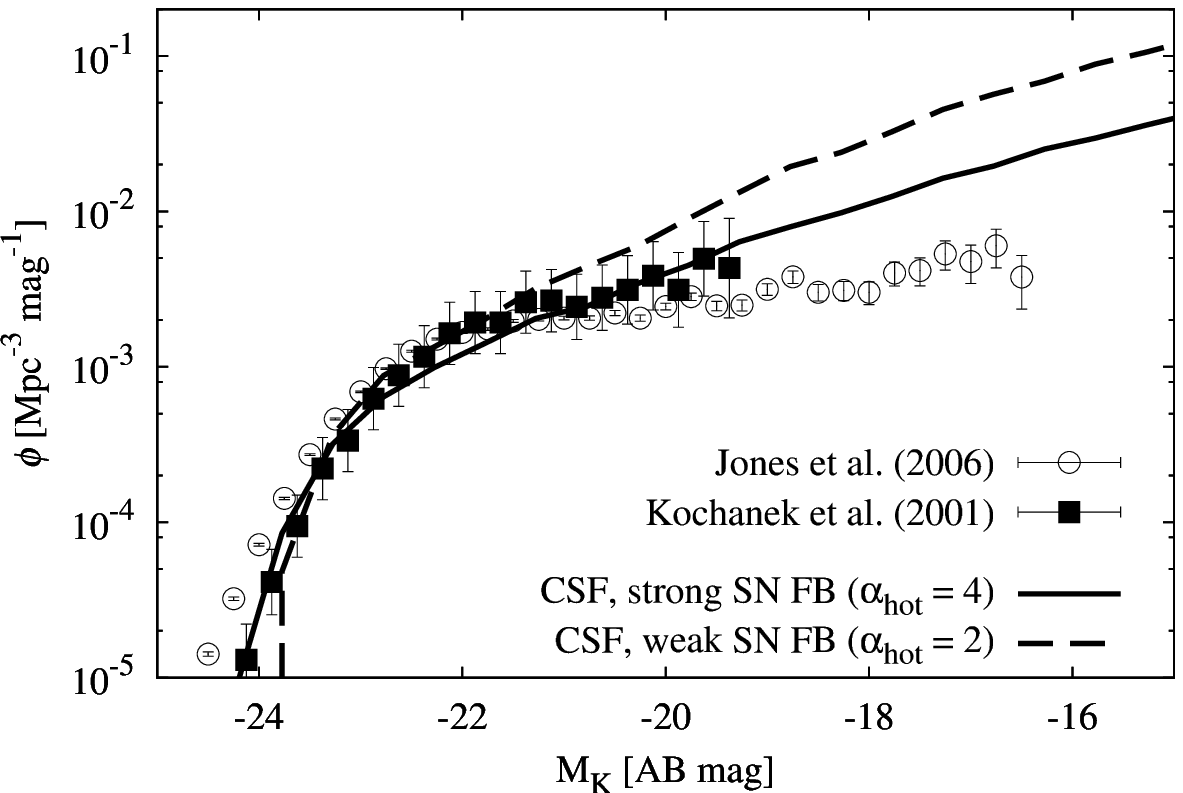} \\
    \end{tabular}
	\caption{ ({\it Left}) The local $g$-band LF compared with the
          NY04 model. The solid and dashed lines represent the NY04
          model with strong SN feedback ($\alpha_{\rm hot} = 4$) and
          weak (reasonable) SN feedback ($\alpha_{\rm hot} = 2$),
          respectively. We only show the results of
          CSF model, since the DSF model gives the almost same
          results. Filled circles indicate the SDSS $g$-band LF
          obtained by Blanton et al. (2005), and open circles are the
          6dF $b_{\rm j}$-band LF obtained by Jones et al. (2006).
          We have transformed the 6dF $b_{\rm j}$-band LF
          to match $g$, by subtracting 0.25 mag (Blanton et al. 2005).
          ({\it Right}) The same as the left panel but for the local
          $K$-band LF.  Data points are the 6dF galaxy survey (Jones et al.
          2006) and 2MASS (Kochanek et al. 2001). } 
\label{fig:NY04_LF_z0p0}
\end{center}
\end{figure*}

\section{Model description}
\label{sec:model}
The detailed description of the basic model, the {\it Mitaka model} is
given by NY04.  Here we focus on the extension made in this work.

\subsection{Star formation recipe}
\label{sec:sf-law}

There are two modes of star formation in our model: quiescent star
formation in galaxy disks and starbursts in major mergers. We follow
the same modeling as NY04 for the starburst mode, where all the cold
gas is converted into stars and hot gas instantaneously.  Since the amount of
stars formed during major mergers is rather minor compared with that in
disk galaxies at low redshift, modeling of the starburst mode does not significantly
change the local luminosity/mass functions.  We change the star formation
recipe for the quiescent mode as follows. The star formation rate is
expressed as
\begin{equation}
\psi = M_{\rm cold} / \tau_{\rm SF}
\end{equation}
where $M_{\rm cold}$ is the cold gas mass, and $\tau_{\rm SF}$ is star
formation time scale.  In the NY04 model, two models for $\tau_{\rm
  SF}$ were considered: constant star formation model (CSF) and
dynamical star formation model (DSF).  In the CSF model, star
formation time scale ($\tau_{\rm SF}$) is constant against redshift,
while in the DSF model $\tau_{\rm SF}$ is proportional to the
dynamical timescale of the host dark matter halo. These models were
expressed as
\begin{eqnarray}
\tau_{\rm SF}=\left\{ \begin{array}{ll}
\tau_{\rm SF}^{0}[1+\beta(V_{\rm circ})] & ({\rm CSF}), \\
\tau_{\rm SF}^{0}[1+\beta(V_{\rm circ})]\left[\cfrac{\tau_{\rm dyn}(z)}{\tau_{\rm dyn}(0)}\right] & ({\rm DSF}), 
\end{array} \right.
\end{eqnarray}
where $\tau_{\rm SF}^{0}$ is a free parameter, $\beta$ is the ratio of
the SF timescale to the reheating timescale by the SN feedback defined
by equation (\ref{eq:SNFB}) (see below), and $\tau_{\rm dyn}(z)$, which is nearly scales as $\propto (1+z)^{-3/2}$, is the dynamical time scale of dark matter halo
at each redshift.

The DSF model is based on an idea that the star formation time scale
is controlled by the dynamical state of an entire galaxy or DM halo,
and star formation activity is highly enhanced at high redshifts
because of the redshift dependence of the dynamical time. 
It is often stated that the AGN feedback is helpful to explain the early 
appearance of massive and quiescent galaxies and to suppress the formation 
of too massive galaxies, but we will later (Section
\ref{sec:LFhighz}) show that enhanced star formation at high redshifts
is also essential, and it is incorporated by DSF in previous models
(e.g. Bower et al. 2006).
\footnote{
Note that Granato et al. (2004) also pointed out the importance of
AGN feedback combined with enhanced star formation at high redshifts
by using a simplified SAM, in a different context of reproducing 
high-z elliptical galaxies rather than solving the problem of formation
of too massive galaxies at the local universe. 
}
However recent observations suggest that the physics
of star formation is determined by the physical state of local
interstellar gas, rather than the dynamical state of entire galaxy
(e.g., Leroy et al. 2008; Lada et al. 2010).  Furthermore, the CSF
model is more favorable than the DSF model to explain the observations
of local dwarf spheroidal galaxies (NY04). In this work we adopt a
star formation law that is determined by the local gas/dust column
density, independent of the galaxy-scale dynamical time.

The local LFs can be reproduced well by both of the
CSF and DSF models of NY04, but the observed cut-off in the $\Sigma_{\rm
  SFR}$--$\Sigma_{\rm gas}$ relation is not reproduced in these
models, indicating a necessity of including another feedback working
at low gas surface density.  Following the discussion in Section
\ref{sec:intro}, we introduce the radiative feedback
depending on dust surface density by adopting the following form of
star formation efficiency (SFE, $\varepsilon \equiv 1/\tau_{\rm SF}$),
\begin{equation}
\varepsilon = \varepsilon_{\rm max} 
\exp(-\tau_{\rm d, th}/\tau_{\rm dust}) + \varepsilon_{\rm min}, 
\end{equation}
where $\tau_{\rm dust}$ is the wavelength-averaged dust opacity. 
In the limit of high dust surface density, SFE becomes constant at
$\epsilon_{\max}$, i.e., $\Sigma_{\rm SFR} \propto \Sigma_{\rm gas}$,
which is consistent with the observation of nearby starburst galaxies
(Kennicutt 1998; Kennicutt \& Evans 2012).
The parameter $\epsilon_{\min}$ controls the strength of
the feedback below the critical dust opacity $\tau_{\rm d,th}$.  We
assume that the dust mass is proportional to the metal mass in the
cold gas phase, and hence $\tau_{\rm dust}$ is given by
\begin{equation}
\tau_{\rm dust} = \frac{1}{2}\frac{\kappa_{\rm d, eff}M_{\rm dust}}{\pi r_{\rm eff}^2} = 2\times10^{-3} \left[\frac{M_{\rm cold} Z_{\rm cold}/{r_{\rm eff}^2}}{\rm M_{\odot}\;Z_{\odot}\;pc^{-2}}\right],
\end{equation}
where $M_{\rm dust}$ is the interstellar dust mass, 
$\kappa_{\rm d, eff} = 2.1\;{\rm pc^2\;M_{\odot}^{-1}}$ is the frequency-integrated effective dust mass opacity
weighted by the local interstellar radiation field (Totani et al. 2011), 
$r_{\rm eff}$ is the effective radius of a galaxy disk, and $Z_{\rm cold}$ 
is the metallicity of cold gas.
We assume that the solar metallicity gas has local dust-to-gas mass ratio, 
0.006 (Zubko et al. 2004).
We follow the typical prescription of SAMs in our model by assuming 
that the disk size is
proportional to the virial radius of host dark matter halos, and
therefore it nearly scales as $r_{\rm eff} \propto 1/(1+z)$ for a fixed halo
mass. 

We treat $\epsilon_{\max}$ as a constant, but introduce the following
two modelings of $\epsilon_{\min}$ for galaxies that are transparent
to UV radiation.  One is simply to assume that $\epsilon_{\min}$ is
also a universal constant.  We cannot assume $\epsilon_{\min} = 0$ in
this case, because SFR becomes zero in metal-free galaxies and hence
galaxies cannot form in the universe. There is a physical
motivation to expect that $\epsilon_{\min}$ evolves with
metallicity. There are two physical processes that would suppress star
formation when UV radiation field is prevalent throughout a galaxy:
H$_{2}$ dissociation and photoelectric heating by dust grains (Schaye
2004; Krumholz et al. 2008, 2009; McKee \& Krumholz
2010). The H$_2$ dissociation should not depend on metallicity, but
the efficiency of photoelectric heating should become larger with
increasing amount of dust, which is assumed here to be proportional to
the metallicity. If the photoelectric heating is relatively important,
we expect that $\epsilon_{\min}$ decreases with metallicity.
Therefore we consider two models (the constant and evolving
$\epsilon_{\min}$ models, hereafter) for the minimum SFE:
\begin{eqnarray}
\varepsilon_{\rm min}=\left\{ \begin{array}{ll}
\varepsilon_{\rm min}^{0} \;\;  & ({\rm constant}\;\varepsilon_{\rm min}), \\[5pt]
\varepsilon_{\rm min}^{0} \exp(-Z_{\rm cold}/Z_{\rm ch}) & ({\rm evolving}\;\varepsilon_{\rm min}), \\
\end{array} \right.
\end{eqnarray}
where $\varepsilon_{\rm min}^{0}$ and $Z_{\rm ch}$ are constant model
parameters.

\subsection{Supernova and AGN feedback}
\label{sec:sn-agn}
In this section, we describe the model of supernova feedback and
AGN feedback since they are highly relevant to star formation process.

{\bf (i) Supernova feedback.}  Following the original Mitaka model, we
assumed that part of cold gas is reheated and ejected from galaxies as a
consequence of supernova feedback at a rate
\begin{equation}
\dot{M}_{\rm reheat} = \psi \, \beta(V_{\rm circ}), 
\end{equation}
where
\begin{equation}
\beta(V_{\rm circ}) = \left( \frac{V_{\rm circ}}{V_{\rm hot}} \right) ^{-\alpha_{\rm hot}},
\label{eq:SNFB}
\end{equation}
where $\dot{M}_{\rm reheat}$ is reheated gas mass per unit time, 
and $V_{\rm circ}$ is the circular velocity of a DM halo. The free 
parameters $\alpha_{\rm hot}$ and $V_{\rm hot}$ are determined 
by the fits to the local LFs, because the faint-end slope and 
characteristic luminosity of LF are sensitively dependent on 
$\alpha_{\rm hot}$ and $V_{\rm hot}$, respectively.

In our model, reheated materials are assumed to be ejected from a
galactic disk into its hot halo gas, with a kinetic energy production
rate of $\sim \dot{M}_{\rm reheat} V_{\rm wind}^{2}/2$.  It is
reasonable to assume that the velocity is determined by the halo
circular velocity, i.e., $V_{\rm wind} \sim V_{\rm circ}$, and the
energy production rate by the supernova feedback is proportional to
SFR $\psi$.  In this case we expect $\alpha_{\rm hot} \sim 2$.
If the scaling is determined by momentum rather than energy, we expect
$\alpha_{\rm hot} \sim$ 1. However, it has been known that a much
stronger feedback efficiency at low velocities than these reasonable
values is required (i.e., $\alpha_{\rm hot}$ = 3--4; NY04; Bower et
al. 2006) to reproduce the faint-end slope of the local LFs. 
In Fig. \ref{fig:NY04_LF_z0p0}, we show this for
the local $g$- and $K$-band LFs using the NY04 model with the two
different model predictions of $\alpha_{\rm hot}$ = 2 and 4.

As already mentioned above, star formation activity in dwarf galaxies
would be suppressed if we adopt the dust opacity-dependent star
formation recipe. Therefore our new model may reproduce the faint-end
LF slopes with a more reasonable efficiency of supernova
feedback. We adopt a reasonable value of 
$\alpha_{\rm hot} = 2$ for all of our new models presented in our work,
and will show that the new model can indeed reproduce the
observed faint-end LF slopes.

{\bf (ii) AGN feedback.}  In the original Mitaka model, in order to
avoid the formation of extremely massive galaxies the cooling process
is applied only to dark matter halos with circular velocity $V_{\rm
  circ} \leq V_{\rm cut}$, where $V_{\rm cut}$ is a free parameter
which is determined to reproduce the local LFs. In the new model, we
introduce the AGN feedback process to make the bright-end of
luminosity function consistent with observations, following the
formulation of Bower et al. (2006).

In our new model, if the following conditions are satisfied the halo is 
prevented from gas cooling;
\begin{equation}
\alpha_{\rm cool} t_{\rm dyn} < t_{\rm cool}
\label{eq:AGNt}
\end{equation}
and
\begin{equation}
\varepsilon_{\rm SMBH} L_{\rm edd} > L_{\rm cool},
\end{equation}
where $t_{\rm dyn}$ is dynamical time scale of the halo, $t_{\rm
  cool}$ is the time scale of gas cooling, $L_{\rm edd}$ is the
Eddington luminosity of the AGN, $L_{\rm cool}$ is the cooling
luminosity of gas, and $\alpha_{\rm cool}$ and $\varepsilon_{\rm
  SMBH}$ are the free parameters which are tuned to reproduce the
observation. The cooling time and dynamical time are calculated at
cooling radius, which is the radius where cooling time scale is equal
to the age of halo. Since our model does not include the formation and
evolution of supermassive black holes, we simply estimated the black
hole mass from the bulge stellar mass, using the observed bulge
mass--black hole mass relation (Marconi \& Hunt 2003). It is unclear
whether the bulge mass--black hole mass relation evolves with redshift
or not, but no evolution hypothesis is consistent with
observations. The AGN feedbacks are important for relatively low
redshift galaxies satisfying the condition of eq. (\ref{eq:AGNt}), and
the possible evolution of the relation would not have a significant
effect. For the results when a SMBH formation model is incorporated into the original Mitaka model, see Enoki, Nagashima \& Gouda (2003), Enoki et al. (2004) and Enoki \& Nagashima (2007).

The condition of eq (\ref{eq:AGNt}) means that the AGN feedback
works only in quasi-hydrostatically cooling haloes (the so-called
``radio mode'' feedback; Croton et al. 2006). In several SAMs, another
mode of AGN feedback is also considered, namely the ``quasar mode'' or
``bright mode'' feedback (Somerville et al. 2008; Bower et al. 2012). 
This feedback mode is considered to be induced by rapid gas
accretion onto supermassive black holes during the major merger phase.
Our model does not include this feedback mode; however, this feedback
mode is only acting in the starburst phase, and therefore it would not
strongly affect the total star formation history or luminosity/mass
function shapes.  Indeed, Bower et al. (2012) showed that the
quasar-mode feedback has only a modest effect on the shape of the
galaxy stellar mass function.

\subsection{Parameter determination}
\label{sec:params}

In summary, there are four new free parameters related to the feedback
depending on dust opacity ($\varepsilon_{\rm max}$, $\tau_{\rm d,
  th}$, $\varepsilon_{\rm min}^{0}$, and $Z_{\rm ch}$), in addition to
the four supernova and AGN feedback parameters in previous models
($\alpha_{\rm hot}$, $V_{\rm hot}$, $\alpha_{\rm cool}$, and
$\varepsilon_{\rm SMBH}$.)  These parameter values of our two models
(constant and evolving $\epsilon_{\min}$ models) are determined by
fitting to the local LFs with the following procedures.
Throughout this paper, we adopt the Salpeter IMF (Salpeter 1955)
with a mass range of 0.1 -- 60 $M_{\odot}$.
The absolute luminosity and colors of individual galaxies are 
calculated using a population synthesis code by Kodama \& Arimoto (1997),
assuming the Galactic extinction curve.

As mentioned above, we fix the supernova feedback parameters to the
reasonable values of $\alpha_{\rm hot} = 2$ and $V_{\rm hot} = 150\;
{\rm km\;s^{-1}}$.  (The $V_{\rm hot}$ value is the same as that in
NY04.) We then find best-fit values of the new parameters introduced
in this work ($\varepsilon_{\rm max}$, $\tau_{\rm th}$,
$\varepsilon_{\rm min}^{0}$, and $Z_{\rm ch}$) by fitting to the local
LFs in relatively faint luminosity range.  Then the AGN feedback
parameters are determined by fitting the bright-end of LFs;
$\alpha_{\rm cool}$ and $\epsilon_{\rm SMBH}$ control the cut-off
luminosity and the shape of the cut-off, respectively.  For both the
constant and evolving $\epsilon_{\rm min}$ models, we found that the
bright-end of local LFs are well reproduced with $\alpha_{\rm cool} =
2.6$ and $\epsilon_{\rm SMBH} = 1.0$. Theoretically, $\alpha_{\rm
  cool} \sim 1$ and $\epsilon_{\rm SMBH} \leq 1$ are required, and the
adopted parameter values are not unreasonable, considering
uncertainties in detailed physical processes.  The determined
parameters are summarized in Table \ref{tb:parameter}.  All of the
other parameters are fixed at the same value with the NY04 model.

\begin{table*}
\begin{center}
\begin{tabular}{cccc}
\hline
\hline
parameter & description
& constant $\varepsilon_{\rm min}$ & evolving $\varepsilon_{\rm min}$ \\
\hline
$\varepsilon_{\rm max}$ ${\rm [Gyr^{-1}]}$    & maximum star formation efficiency & 10.0 & 10.0 \\
$\tau_{\rm d, th}$                            & threshold dust opacity & 1.0 & 1.0 \\
$\varepsilon^{0}_{\rm min}$ ${\rm [Gyr^{-1}]}$& minimum star formation efficiency & $1.5\times10^{-4}$ & $5.0\times10^{-3}$ \\
$Z_{\rm ch}$ $[Z_{\odot}]$& characteristic metallicity 
for $\varepsilon_{\min}$ evolution & -- & 0.02 \\
$\alpha_{\rm hot}$ & SN feedback controlling parameter 
  & 2.0 (fixed) & 2.0 (fixed) \\
$V_{\rm hot} [{\rm km s^{-1}}]$ &  SN feedback controlling parameter 
  & 150 (fixed) & 150 (fixed)\\
$\alpha_{\rm cool}$ & AGN feedback controlling parameter & 2.6 & 2.6 \\
$\varepsilon_{\rm SMBH}$ & AGN feedback controlling parameter & 1.0 & 1.0 \\
\hline
\end{tabular}
\caption{The model parameters in the constant $\varepsilon_{\rm min}$
  and evolving $\varepsilon_{\rm min}$ model. All of the other
  parameters are fixed at the same value with the NY04 model. See
  Section \ref{sec:sf-law} and \ref{sec:sn-agn} for parameter
  descriptions and Section \ref{sec:params} for the parameter
  determination procedures.  }
\label{tb:parameter}
\end{center}
\end{table*}

\section{Results}
\label{sec:Results}
\subsection{local luminosity functions}
\begin{figure*}
  \begin{center}
  	\begin{tabular}{cc}
		\includegraphics[width=85mm]{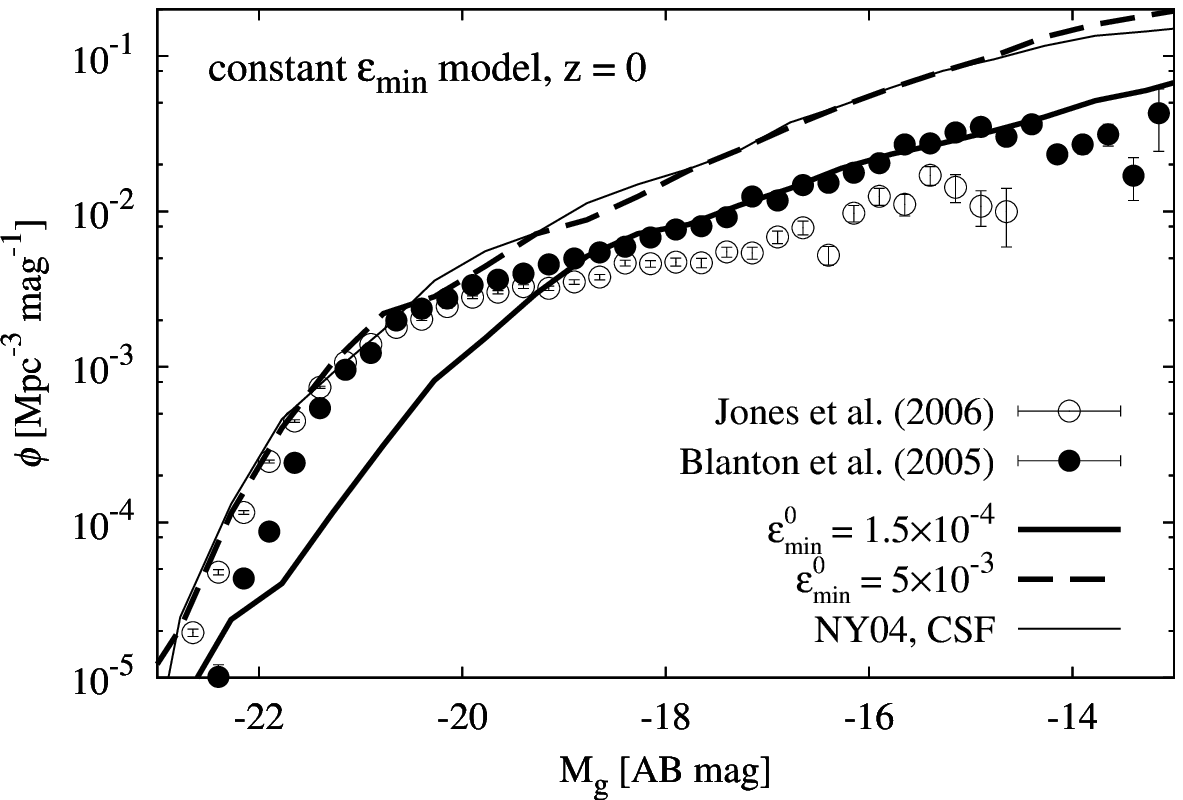} &
		\includegraphics[width=85mm]{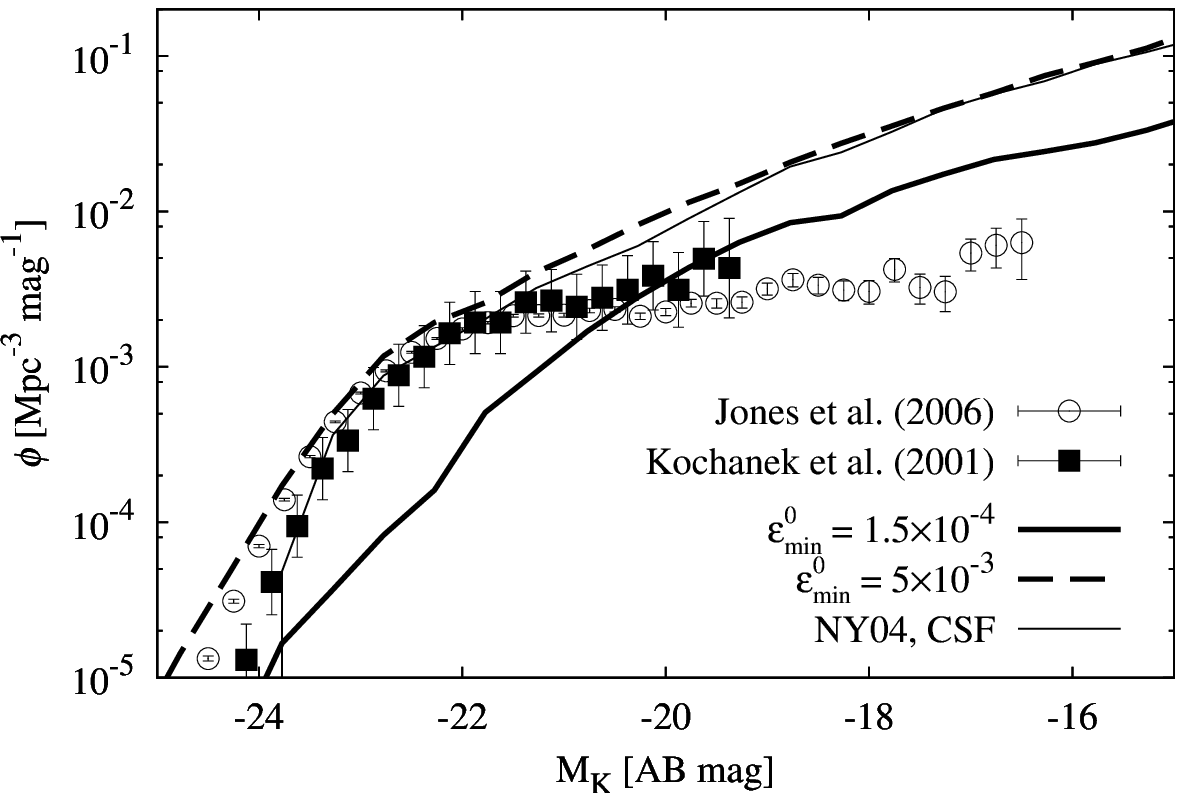} \\\
    \end{tabular}
	\caption{Local $g$- ({\it Left}) and $K$-band ({\it Right}) LFs
          for the constant $\varepsilon_{\rm min}$ model.  Data points
          are the same as Fig. \ref{fig:NY04_LF_z0p0}.  The thick
          solid line represents the result with the adopted parameter
          values listed in Table \ref{tb:parameter}. We also plotted
          the result of NY04 with CSF model for
          comparison (thin solid line).  The dashed line shows the
          same model but with a different value of
          $\varepsilon^{0}_{\rm min} = 5\times10^{-3}$ ${\rm
            Gyr^{-1}}$.
          The weak SN feedback mode ($\alpha_{\rm hot}$ = 2) is adopted in all models.
            }
\label{fig:g_LF_z0p0_model1}
\end{center}
\end{figure*}

In Fig. \ref{fig:g_LF_z0p0_model1}, we show the local $g$-
and $K$-band LFs for the constant $\varepsilon_{\rm min}$ model. The
result of NY04 model with CSF model and weak SN feedback (i.e.,
$\alpha_{\rm hot}$ = 2.0) is also shown for comparison. Since there is
not much differences between the results of CSF and DSF model at the
local universe, we only plot the result of CSF model. The data points
are the SDSS, 6dF, and 2MASS measurements of the local LFs (Blanton et
al. 2005; Jones et al. 2006; Kochanek et al. 2001).  We have transformed the 6dF $b_{\rm
  j}$-band LF to match $g$-band LF, by subtracting 0.25 mag (Blanton et
al. 2005). It can be seen that the faint-end slope of LF obtained by
Jones et al. (2006) is flatter than that obtained by Blanton et
al. (2005).  One of the reasons of this discrepancy would be a local
fluctuation of galaxy abundances.  The data of Blanton et al. (2005)
is deduced from deeper but narrower survey, while the data of Jones et
al. (2006) is based on the shallower but wider surveys.

Two new model curves with the different values of $\varepsilon_{\rm
  min}^{0} = 1.5\times10^{-4}$ and $5\times10^{-3}$ are also shown, 
and it can be seen that the change of $\varepsilon_{\rm min}^{0}$ results in just a
change of normalization of LF, keeping the LF shape roughly unchanged;
steeper faint end of the model compared with the data still remains.
Since the constant $\varepsilon_{\rm min}$ model cannot reproduce
the local LFs, we will focus on the evolving $\varepsilon_{\rm min}$
in the following of this paper.

\begin{figure*}
  \begin{center}
    \begin{tabular}{cc}
	\includegraphics[width=85mm]{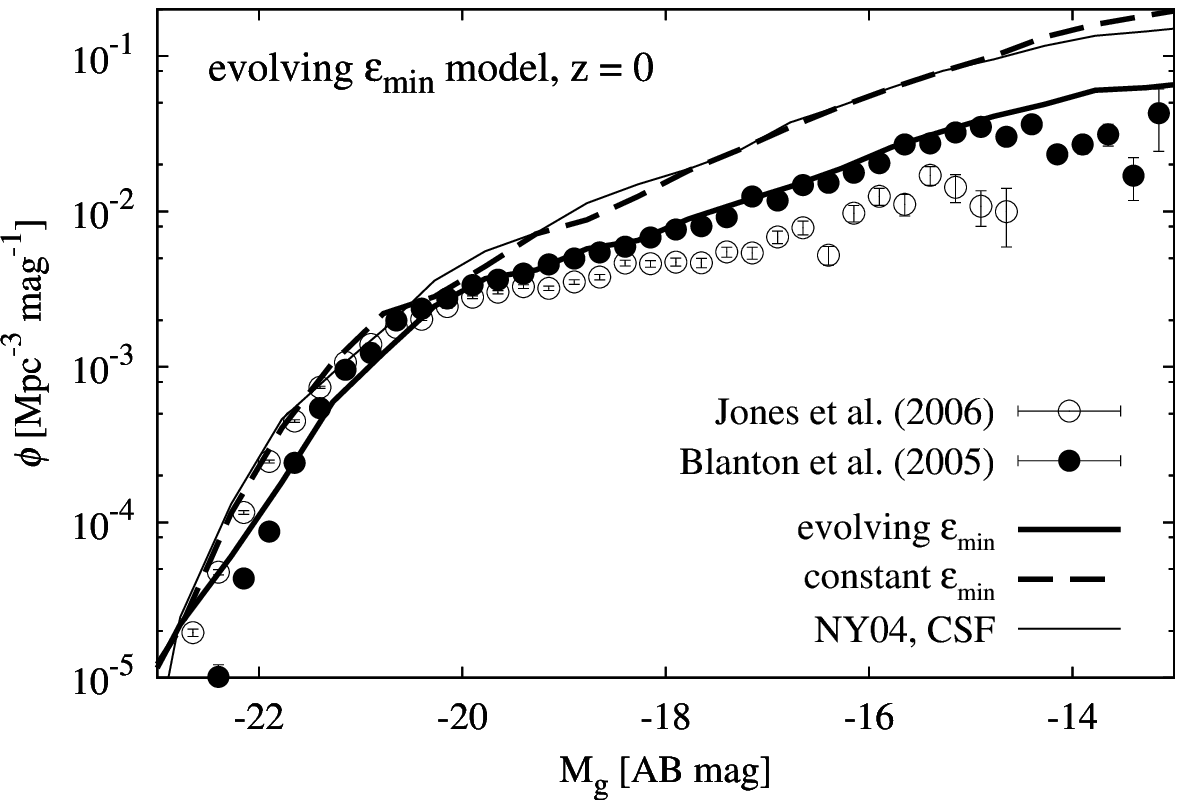} &
	\includegraphics[width=85mm]{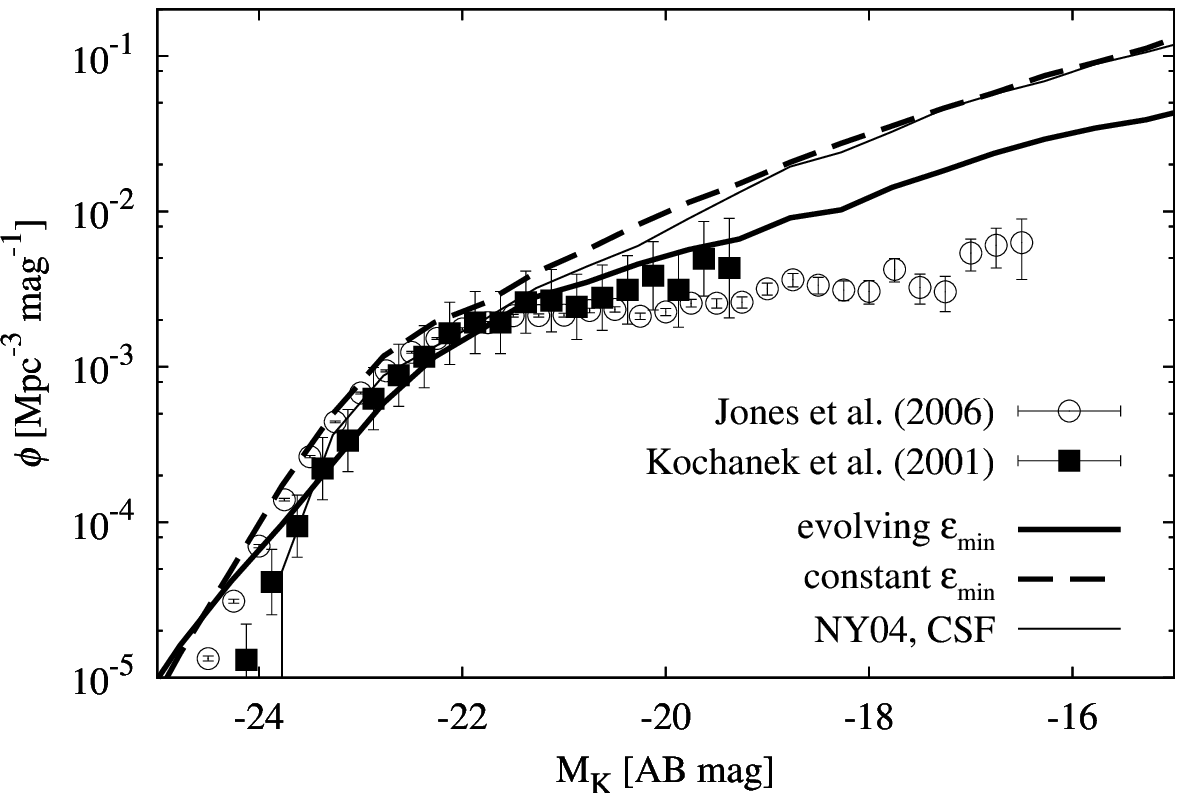} \\\
    \end{tabular}
	\caption{Local g- ({\it Left}) and $K$-band ({\it Right}) LFs
          for the evolving $\varepsilon_{\rm min}$ model.  Data points
          are the same as Fig. \ref{fig:NY04_LF_z0p0}.  The thick
          solid line represents the result with the adopted parameter
          values listed in Table \ref{tb:parameter}.  
          The dashed line represents
          the result of constant $\varepsilon_{\rm min}$ model, with
          the same value of $\varepsilon^{0}_{\rm min}$ as the
          evolving $\varepsilon_{\rm min}$ model,
          $5.0\times10^{-3}$; therefore the difference of the evolving
          and constant $\varepsilon_{\rm min}$ in this figure is only
          due to the metallicity dependence of the minimum SFE,
          $\varepsilon_{\rm min}$. 
           We also plotted the NY04 model for
          comparison (thin solid line).
          The weak SN feedback ($\alpha_{\rm hot}$ = 2) is adopted in all models.
          }
\label{fig:g_LF_z0p0_model2}
\end{center}
\end{figure*}

In Fig. \ref{fig:g_LF_z0p0_model2}, we show the local $g$- and
$K$-band LFs for the evolving $\varepsilon_{\rm min}$ model. The
results of the constant $\varepsilon_{\rm min}$ model and the NY04
model with weak SN feedback are also plotted for comparison.  In the
evolving $\varepsilon_{\rm min}$ model, the formation of dwarf
galaxies are significantly suppressed and the model well reproduces
the observed LFs at overall magnitudes. Note that we used the same
value of $\varepsilon_{\rm min}^{0}$, $5\times10^{-3}$, for the
constant $\varepsilon_{\rm min}$ and evolving $\varepsilon_{\rm min}$
models in this plot, and therefore the difference of two models are
only due to the metallicity dependence of $\varepsilon_{\rm min}$.

The LF faint end is suppressed in the evolving $\varepsilon_{\rm min}$
model because the star formation in small galaxies at $z \sim 0$ is
suppressed by the feedback introduced to the model. This feedback is
stronger at smaller galaxies by the condition for dust opacity,
because more massive galaxies generally have higher metallicity and
higher mass surface density when the ratio of gas mass to dark matter
is fixed ($\Sigma_{\rm DM} \propto M^{1/3}$ at a fixed redshift).
However, the success of the evolving $\varepsilon_{\rm min}$ model
against the constant model indicates that the feedback depending only
on dust opacity is not sufficient. In such a model, the number of
massive galaxies is also reduced when the feedback is strong enough to
suppress the LF faint-end, as seen in Fig. \ref{fig:g_LF_z0p0_model1}.
This is because star formation in the early phase of massive galaxies
is suppressed by low dust opacity when their metallicity is still
low. Therefore another dependence of the feedback on metallicity,
which is motivated by the dust photoelectric heating process, is
essential to allow formation of massive galaxies at $z \sim 0$.

\subsection{luminosity function at high redshift}
\label{sec:LFhighz}

\begin{figure*}
  \begin{center}
  	\includegraphics[width=150mm]{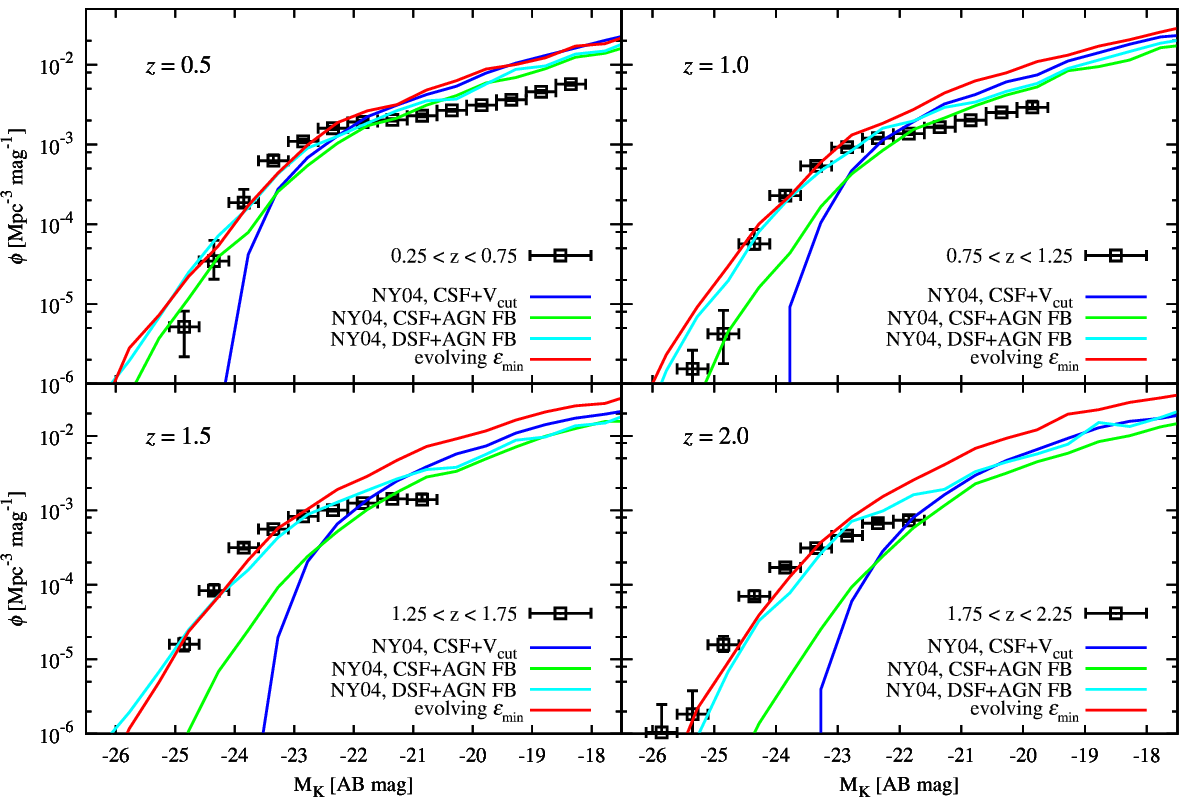}
	\caption{The evolution of $K$-band LFs at $z$ = $0.5, 1.0, 1.5$
          and 2.0.  The solid lines represent the results of NY04 with
          CSF and $V_{\rm cut}$ model (blue), NY04 with CSF and AGN
          feedback model (green), NY04 with DSF and AGN feedback model
          (cyan), and our new model (the evolving $\varepsilon_{\rm
            min}$ model with AGN feedback, red).  Open squares are the
          observed data obtained by Cirasuolo et al. (2010).
          In the evolving $\varepsilon_{\rm min}$ model, weak SN
feedback model ($\alpha_{\rm hot}$ = 2) is adopted, while in the other
models adopted strong SN feedback model ($\alpha_{\rm hot} = 4$).  The
parameters of AGN feedback model are fixed as the same value in all
models.}
\label{fig:K-evo}
\end{center}
\end{figure*}

In Fig. \ref{fig:K-evo}, we show the $K$-band LFs at $z$ = 0.5, 1,
1.5, and 2 for the evolving $\varepsilon_{\rm min}$ model, in
comparison with the observed data of Cirasuolo et al. (2010).  To see
the effect of star formation recipe and AGN feedback, we also show
some variations of NY04 models: CSF with $V_{\rm cut}$ model, CSF with
AGN feedback model, and DSF with AGN feedback model.
In the evolving $\varepsilon_{\rm min}$ model, weak SN
feedback model ($\alpha_{\rm hot}$ = 2) is adopted, while in the other
models adopted strong SN feedback model ($\alpha_{\rm hot} = 4$).  The
parameters of AGN feedback model are fixed as the same value in all
models.

It can be seen that the CSF with $V_{\rm cut}$ model significantly
underestimates the bright end of LFs, especially at high redshift. If
we introduced AGN feedback into the CSF model, the situation is
slightly improved since AGN feedback does not efficiently work at high
redshift; however, the model still underestimates the bright-end of
LFs. By contrast, the DSF + AGN feedback model well reproduces the
observations at all redshift range. This is because the DSF model has
shorter star formation time scale than CSF model at high redshift. 

It has been thought that the AGN feedback plays an important role in
reproducing the downsizing trend of cosmological galaxy formation
(e.g., Bower et al. 2006); however, these results suggest that the
dependence of star formation time scale on the halo or galaxy
dynamical scale is also essential, as well as the AGN feedback.  In
most of SAMs, star formation time scale is simply proportional to the
dynamical time scale of host DM halo or galaxy disk (e.g., Cole et
al. 2000; NY04; Bower et al. 2006).  However, recent observations
suggest that star formation time scale is seems to be determined by
local physical condition in a galaxy, rather than the dynamical time
scale of an entire galaxy (see section \ref{sec:intro}).

By contrast, our new model successfully reproduces the high-$z$
$K$-band LFs, without introducing the dependence of star formation on
the dynamical time scale of a DM halo or galaxy.  Star formation time
scale is shorter in massive galaxies at higher redshift also in our
new model, but it is because of the newly introduced feedback
depending on metallicity and dust opacity, and the general trend that
high redshift star-forming massive galaxies have high dust opacity. It
should be noted that the baseline star formation time scale
$\epsilon_{\max}$, which determines star formation rate when the
feedback is not effective, is a universal constant in our model.

Our models overestimate the abundance of dwarf galaxies, especially at
high redshift. This is not only for the new feedback model, but also
for the conventional models with the AGN feedback.  It might suggest
that there are some missing physical processes in the presented
models; however, there may also be a large uncertainty on the
measurement of the faint-end high-$z$ $K$-band LFs, by e.g., detection
efficiency around the detection limit, errors on determination of the
rest-frame luminosities, or cosmic variance. Therefore we do not
discuss this issue further in this paper.

\subsection{The cosmic star formation history}
\begin{figure*}
  \begin{center}
  	\begin{tabular}{cc}
        	\includegraphics[width=85mm]{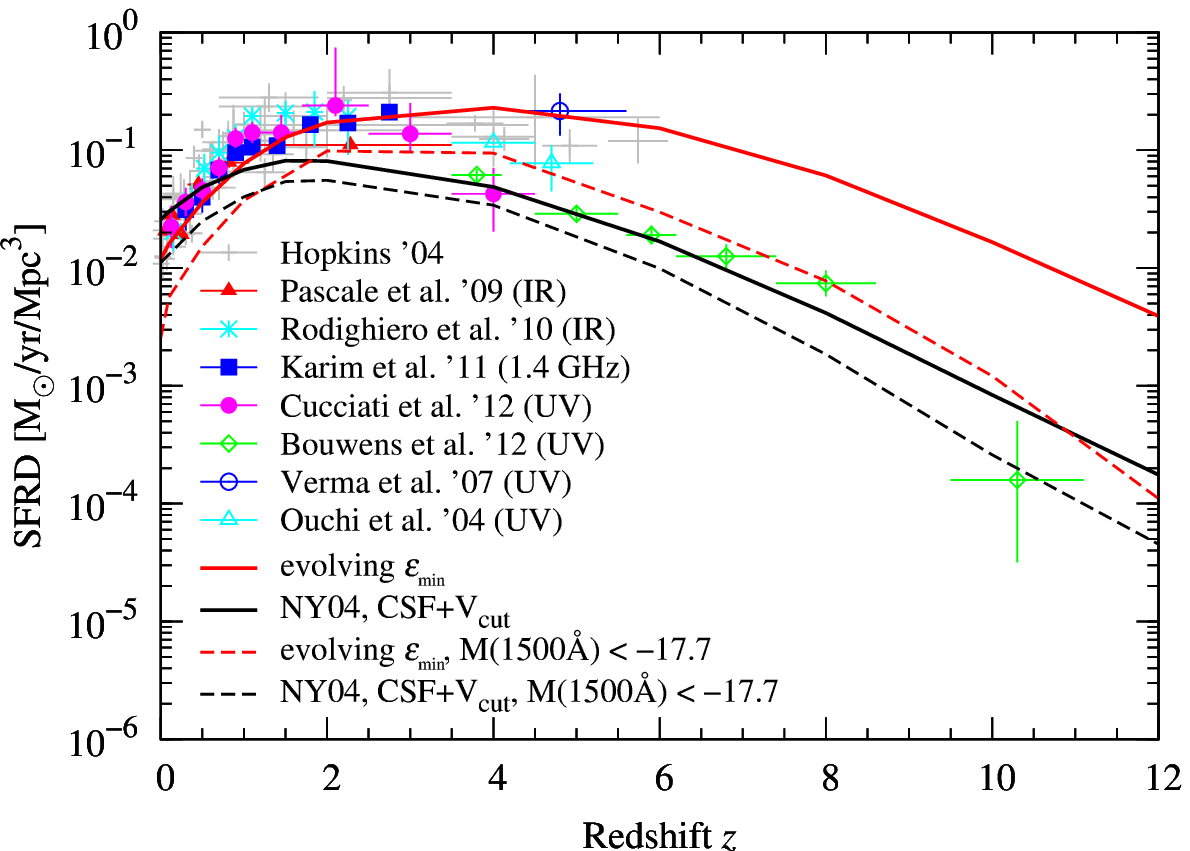} &
		\includegraphics[width=85mm]{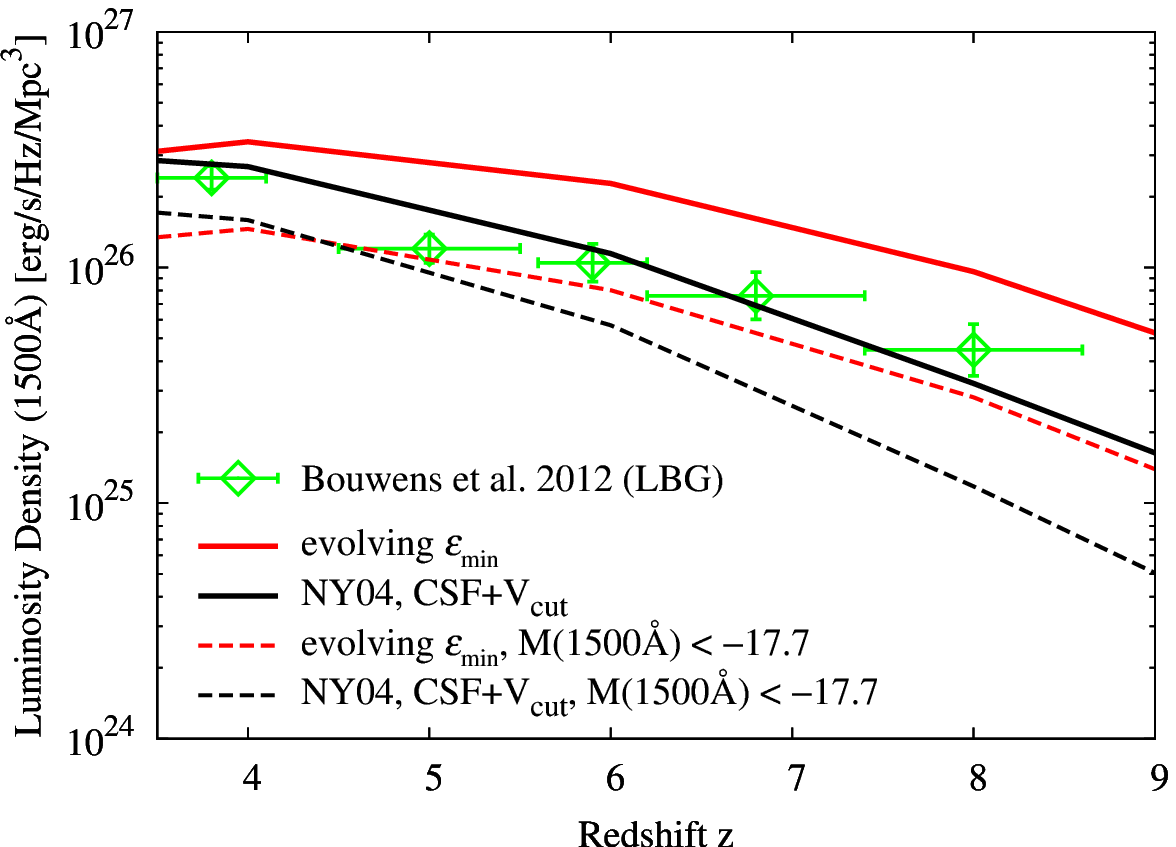} \\
    \end{tabular}
	\caption{ {\it(Left)} The cosmic SFR density
          evolution. The solid lines show the total SFR (i.e.,
          integrated over all luminosity range) per unit comoving
          volume in the evolving $\varepsilon_{\rm min}$ model (red)
          and the NY04 with CSF and $V_{\rm cut}$ model (black). 
          The dashed red and black lines are the same as the solid
          lines, but integrated only for galaxies brighter than
          $M_{\rm AB}(1500 {\rm \AA}) < -17.7$ (extinction uncorrected
          magnitude).  We also plot the observed data estimated by
          dust continuum emission from FIR to radio band (Pascale et
          al. 2009; Rodighiero et al. 2010; Karim et al. 2011) and UV
          continuum (Cucciati et al. 2012; Bouwens et al. 2012; 
          Verma et al. 2007; Ouchi et al. 2004). 
          The data points of Hopkins (2004) are the compilation of
          observations in several wavelengths and methods.  All the
          data points are corrected for extinction, by the methods
          adopted in individual references.  The open symbols for
          UV continuum-based estimates at $z > 4$ are obtained by integrating
          LF down to the limiting magnitudes of each survey; the
          limiting magnitude of $M_{\rm AB}(1500 {\rm \AA}) < -17.7$
          adopted by Bouwens et al. (2012) is the same as that for the
          dashed model curves. The other filled symbol data points are
          integration of LFs in the entire magnitude range. 
          {\it(Right)} The redshift evolution of luminosity density at
          rest-frame 1500 \AA, without correction about extinction. 
          The model curves are the same as
          the left panel.  The data points are integrations in the
          range of $M(1500 {\rm \AA}) < -17.7$. 
          }
\label{fig:csfh}
\end{center}
\end{figure*}

In Fig. \ref{fig:csfh}, we compare the cosmic star formation history
(i.e., SFR per unit comoving volume as a function of redshift)
of our theoretical models with the observed data.  In the new evolving
$\varepsilon_{\rm min}$ model, star formation activity is
significantly enhanced at high redshifts, and it becomes about an
order of magnitude higher than the old NY04 model with CSF and $V_{\rm
  cut}$ at $z \gtrsim 6$. This enhancement is caused by galaxies having
high dust opacity or low metallicity in which the feedback is not
strongly working.

However, the difference between the new model and NY04 is rather
modest when galaxies are limited into those with $M_{\rm UV}(1500$
${\rm \AA}) < -17.7$. This is because the enhancement of SFR in the
new model is mainly by dusty galaxies, and such galaxies are faint in
UV.  Even if UV luminosity is brighter than the observational limiting
magnitude, dusty and hence red galaxies may be missed in the selection 
criteria of Lyman break galaxies (Bouwens et al. 2012).  
As a result, both models
are roughly consistent with the observed data when the limiting
magnitudes are appropriately taken into account, also considering
various sources of uncertainties in the estimation of cosmic SFR
density, such as the faint-end slope of the LF, correction of dust
extinction, contamination from old stellar populations to the IR
luminosity, assumed stellar spectra and IMF.  Recently, Kobayashi et
al. (2013) have shown that a discrepancy by a factor of 2--3 can
indeed arise from overcorrection for dust obscuration and
luminosity-to-SFR conversion.

Comparison in the rest-frame UV luminosity density would suffer from
less uncertainties than that in SFR density, and this is shown in the
right panel of Fig. \ref{fig:csfh}.  Interestingly, the new model
gives a quantitatively better fit to the data than the old NY04 model,
though the discrepancy between the NY04 model and the data may still
be within the systematic uncertainties.  The new model shows a flatter
evolutionary trend toward higher redshift than the NY04 model, which is
also in good agreement with the data.

It would be interesting to search for the UV-faint, dusty star-forming
galaxies at high redshifts predicted by the new model, by future observations in other
wavelengths, e.g., submillimeter surveys by ALMA.
They are below the magnitude limit in the current surveys in UV 
but significantly contributing to the total cosmic SFR.

\section{Radiative Feedback Depending on Gas Surface Density}
\label{sec:gasmodel}
In this paper we have examined a new feedback process depending on
galaxy-scale dust surface density. Although observations and
theoretical considerations suggest that a dust surface density plays
an important role in determining the galaxy-scale star formation rate,
the original Kennicutt-Schmidt relation is the scaling relation
between SFR surface density and gas surface density, not dust surface
density. Therefore it is interesting to compare our new model with
another one assuming a star formation law depending on gas surface
density, and examine whether the dust opacity dependence is essential
or not in our new model.

Here we adopt the following simple formula of SFE,
\begin{equation}
\varepsilon = \varepsilon_{\rm max} \, 
\exp(-\Sigma_{\rm gas, th}/\Sigma_{\rm gas}) \ ,
\end{equation}
where $\Sigma_{\rm gas} = M_{\rm cold}/\pi r^{2}$ is the gas surface
density, and $\Sigma_{\rm gas, th}$ is the threshold of gas surface
density below which SFE rapidly decreases. In what follows we will
refer to this model as ``the $\Sigma_{\rm gas}$ model''. In the
$\Sigma_{\rm gas}$ model we do not introduce the lower limit of SFE,
$\varepsilon_{\rm min}$, since $\varepsilon$ has a finite value in
this model even in galaxies without any metal or dust, provided that
$\Sigma_{\rm gas}$ is higher than the threshold value.

\begin{figure*}
  \begin{center}
    \begin{tabular}{cc}
	\includegraphics[width=85mm]{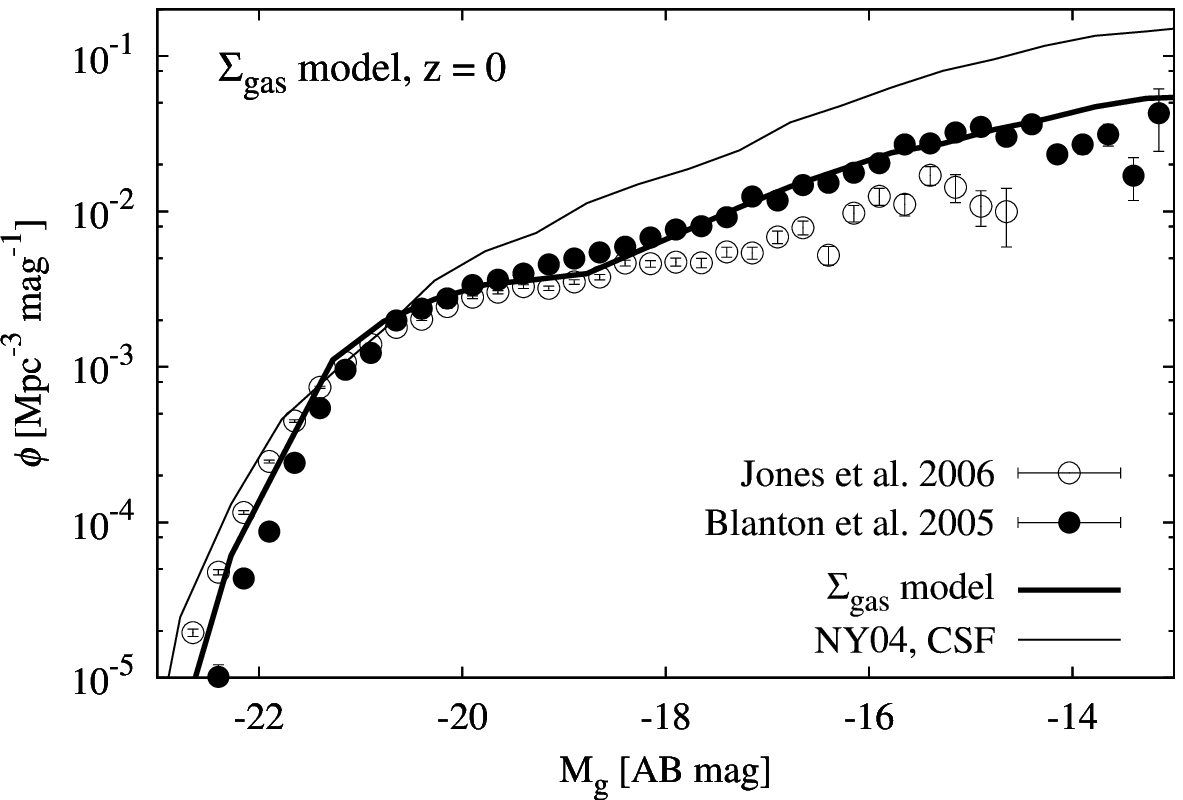} &
	\includegraphics[width=85mm]{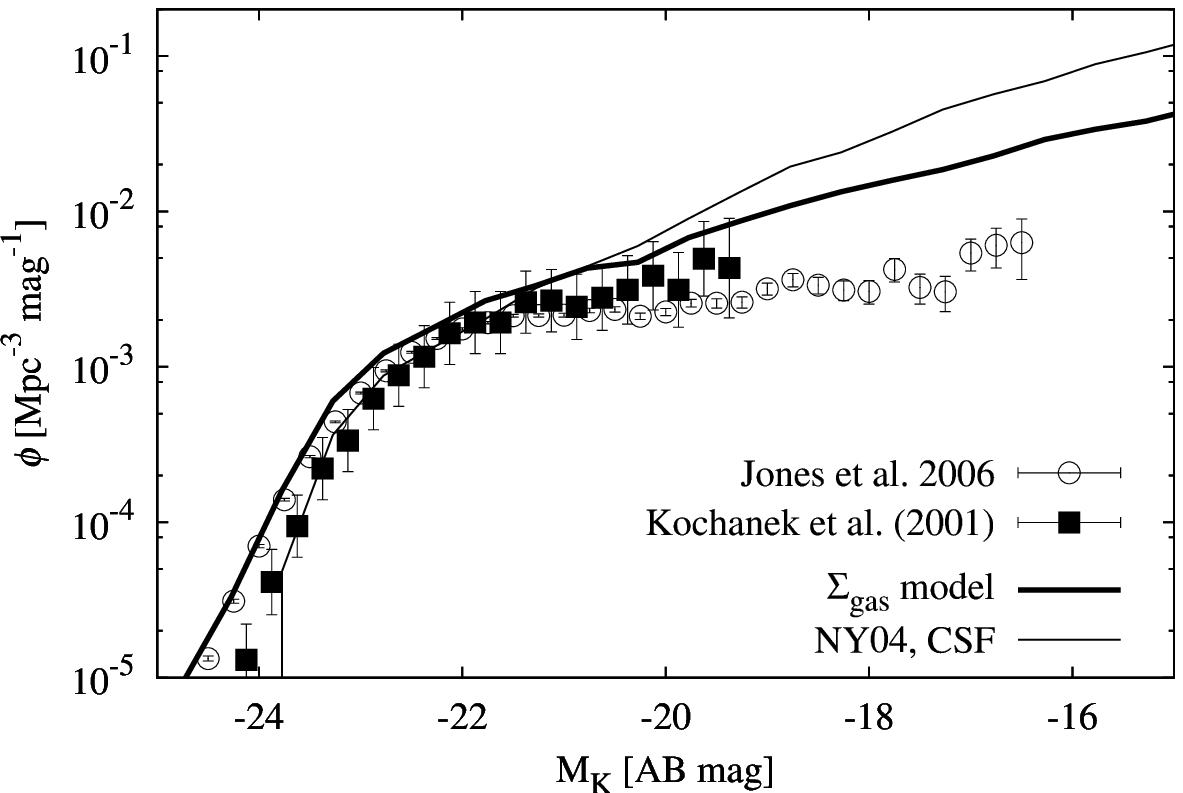} \\
    \end{tabular}
	\caption{Local g- ({\it Left}) and $K$-band ({\it Right}) LFs
          for the $\Sigma_{\rm gas}$ model (thick solid line).  We
          also plotted the results of the NY04 model (thin solid line) for
          comparison.  The weak SN feedback mode ($\alpha_{\rm hot} = 2$) is adopted in all models. 
          The observed data points are the
          same as Fig. \ref{fig:NY04_LF_z0p0}.  }
\label{fig:LF_z0p0_model3}
\end{center}
\end{figure*}

In Fig. \ref{fig:LF_z0p0_model3}, we show the local $g$- and
$K$-band LFs for the $\Sigma_{\rm gas}$ model. We also show the result
of NY04 model (CSF and weak SN feedback is adopted) for comparison.
The adopted parameters are $\varepsilon_{\rm max} = 10$ ${\rm
  Gyr^{-1}}$ and $\Sigma_{\rm gas, th} = 50\;M_{\odot}\;{\rm pc^{-2}}$.
This $\Sigma_{\rm gas, th}$ roughly corresponds to $\tau_{\rm dust}
\sim 0.3$ when $Z \sim Z_{\odot}$. In this model, we also adopted the
weak SN feedback parameter ($\alpha_{\rm hot} = 2$). Other parameters
are fixed at the same with the adopted values of the evolving
$\varepsilon_{\rm min}$ model (see Table
\ref{tb:parameter}). Procedures of the parameter determination is the
same with the dust-opacity dependent feedback models (see Section
\ref{sec:params}).  We can see that the formation of dwarf galaxies is
significantly suppressed, and the $\Sigma_{\rm gas}$ model also well
reproduces the observed LFs. Thus the dependence on dust opacity or
gas surface density cannot be discriminated only in local LFs.

\begin{figure*}
  \begin{center}
  	\includegraphics[width=150mm]{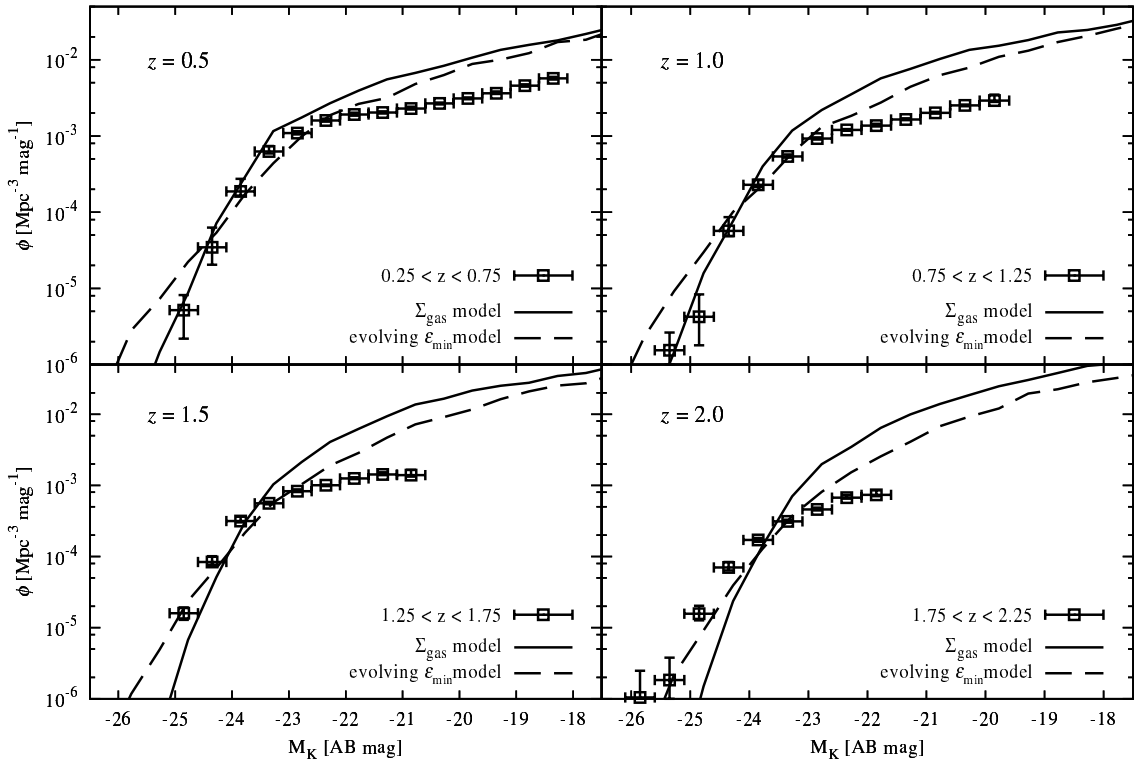}
    	\caption{The evolution of $K$-band LFs at $z = 0.5, 1.0, 1.5$
          and 2.0 for the $\Sigma_{\rm gas}$ model.  We also plotted
          the results of the evolving $\varepsilon_{\rm min}$ model
          for comparison.  Open squares are the observed LFs obtained
          by Cirasuolo et al. (2010). 
            }
\label{fig:K-evo2}
\end{center}
\end{figure*}

However, they show different redshift evolution of $K$-band LFs 
as shown in Fig. \ref{fig:K-evo2}. In this figure we also show 
the results of the evolving $\varepsilon_{\rm min}$ model for comparison. 
It can be seen that the $\Sigma_{\rm gas}$ model predicts more dwarf galaxies
and less massive galaxies than the evolving $\varepsilon_{\rm min}$
model, especially at high redshift.  This difference can be explained
as follows. There is a well-known trend of higher metallicity for more
massive galaxies, i.e., the so-called stellar mass--metallicty
relation (e.g., Tremonti et al. 2004). Therefore the model
depending on dust opacity should have a stronger trend of higher star
formation efficiency for more massive galaxies than the $\Sigma_{\rm
  gas}$ model at a fixed redshift.  The $\Sigma_{\rm gas}$ model
predicts high star formation efficiency for dwarf galaxies at high
redshifts because of high gas density, and the result of Fig.
\ref{fig:K-evo2} indicates that the predicted efficiency is too high
compared with observations. The new model presented here depending on
dust opacity gives a better fit about this observation.

\section{Summary}
\label{sec:summary}

In this paper, we have considered a new feedback mechanism on star
formation depending on galaxy-scale mean optical depth to absorption
by dust grains, and examined the effect on galaxy luminosity functions
and their cosmological evolution, making use of a semi-analytic model
of galaxy formation.  The introduction of such feedback process is
motivated not only by theoretical considerations but also by recent
observations, which indicate that star formation activity is
significantly suppressed in galaxies that are transparent to UV
radiation.  The structure formation theory predicts that the
dust-opacity becomes higher in massive objects and at higher redshifts
for a fixed dust-to-gas ratio; therefore it is expected that the
faint-end of local LFs would be suppressed, which is required for the
current galaxy formation models to match the observations. Note that extremely
strong supernova feedback was required in the conventional models to
reproduce the observed faint end of local LFs. Such feedback process
would also accelerate the formation of massive galaxies at high
redshifts.

We have tested a few models about star formation feedback, and the best fit
with observations is found with the model in which star formation is
suppressed when the galaxy-scale dust opacity is low and metallicity
is higher than a critical value (the evolving $\varepsilon_{\min}$
model).  The latter condition is introduced phenomenologically, but
theoretically motivated by the process of photoelectric heating by
dust grains. In this model formation of dwarf galaxies at $z \sim 0$
is significantly suppressed, and the model successfully reproduces the
faint-end slope of local LFs with a physically natural strength of the
SN feedback.

The new model also succeeded in reproducing the number density of
high-$z$ massive galaxies.  The early appearance of massive galaxies
have been explained by the AGN feedback process; however we have found
that the star formation model is also important as well as the AGN
feedback.  In most of SAMs, star formation time scale is assumed to be
proportional to the dynamical time scale of a host DM halo or galaxy
disk (e.g., Cole et al. 2000; Bower et al. 2006; NY04). This is
essential to explain the early appearance of massive galaxies, because
the model with a constant star formation time scale cannot reproduce
it even if the AGN feedback is incorporated.  However, recent
observations suggest that the star formation efficiency is closely
related to the gas or dust surface density, rather than the dynamical
time scale of an entire galaxy or halo (see section
\ref{sec:intro}). Our new model incorporating the AGN feedback can
explain the number density of high-$z$ massive galaxies with the
observationally suggested star formation law.
The new model is also consistent with the observed cosmic star formation
history. 

We also tested a star formation feedback model depending simply on the
gas surface density (the $\Sigma_{\rm gas}$ model), rather than the
dust opacity, to examine whether the dust opacity is essential or not.
Although this model can also reproduce the shape of the local LFs, the
difference from the evolving $\varepsilon_{\min}$ model appears in the
mass function (or $K$-band LF) at high redshifts.  The evolving
$\varepsilon_{\min}$ model predicts more galaxies than the 
$\Sigma_{\rm gas}$ model at the bright end of $K$-band LFs at $z \sim
2$, which is in better agreement with the observed data.

To conclude, we have found that the feedback depending on galaxy-scale
dust opacity has significant effects on the cosmological galaxy
formation, and has good properties to solve some of the problems found
in the previous theoretical models.  However, it should also be noted
that there are still various uncertainties in our model.  For example,
we determined the value of star formation efficiency under the dust
opacity threshold phenomenologically from fits to the luminosity
function data, but these results should be examined in light of
theoretical studies of star formation.  We assumed that dust mass is
simply proportional to the metal mass, but it is not obvious that this
proportionality is valid for all galaxies. More observational and
theoretical studies on formation/evolution of dust grains are
desirable to establish a better star formation modeling for
cosmological galaxy formation.

\section*{Acknowledgments}
RM has been supported by the Grant-in-Aid for JSPS Fellows.
MN and TTT have been supported by the Grant-in-Aid for the Scientific
Research Fund (25287041 for MN, 23340046 and 24111707 for TT) 
commissioned by the Ministry of Education, Culture, Sports, 
Science and Technology (MEXT) of Japan.
TTT also has been partially supported from the Strategic Young
Researches Overseas Visits Program for Accelerating Brain
Circulation  from the MEXT.

\end{document}